# The Launch Dynamics of Supersonic Projectiles

C. K. Muthukumaran[1]
*Department of Aerospace, Indian Institute of Space Science and Technology, Trivandrum, Kerala-695547*

G. Rajesh[2†]
*Dept. of Mechanical and Automotive Engineering, Keimyung University, 1095 Shindang Dong, Dalseo-Gu, Daegu-704-701, SOUTH KOREA*
*rajesh@kmu.ac.kr*

H. D. Kim[3]
*School of Mechanical Engineering, Andong National University, SOUTH KOREA,*

**The aerodynamics of projectiles launched from barrels of various devices is quite complicated due to its interactions with the unsteady flow field around it. A computational study using a moving grid method is performed here to analyze various fluid dynamic phenomena in the near field of a gun, such as, the projectile-shock wave interactions and interactions between the flow structures and the aerodynamic characteristics of the projectile when it passes through various flow interfaces. Cylindrical and conical projectiles have been employed to study such interactions and the fluid dynamics of the flow fields. The aerodynamic characteristics of the projectile are hardly affected by the projectile configuration during the process of the projectile overtaking the primary blast wave for small Mach numbers. However, it is noticed that the projectile configurations do affect the unsteady flow structures before overtaking and hence the unsteady drag coefficient for the conical projectile shows considerable variation from that of the cylindrical projectile. The projectile aerodynamic characteristics during the interaction with the secondary shock wave are also analyzed in detail. It is observed that the change in the characteristics of the secondary shock wave during the interaction is fundamentally different for different projectile configurations. Both inviscid and viscous simulations were carried out to study the projectile aerodynamics and the fluid dynamics. Though the effect of the viscosity on the projectile aerodynamic characteristics is not significant, the viscosity greatly affects the unsteady flow structures around the projectile.**

___________________________

[1] *Research Scholar,* Department of Aerospace, Indian Institute of Space Science and Technology, Trivandrum, Kerala-695547

[2†] *Corresponding Author,* Assistant Professor, Dept. of Mechanical and Automotive Engineering, Keimyung University, 1095 Dalgubeoldaero Dalseo-Gu, Daegu 704-701, SOUTH KOREA.

[3] Professor, School of Mechanical Engineering, Andong National University, SOUTH KOREA,

## Nomenclature

| | | |
|---|---|---|
| $A$ | = | Projected frontal area of the projectiles. |
| $a$ | = | Speed of sound, m/s. |
| $C_d$ | = | Coefficient of drag. |
| $D$ | = | Drag force, N. |
| $M$ | = | Mach number. |
| $M_{p1}$ | = | Projectile Mach number relative to still air. |
| $M_{p2}$ | = | Projectile Mach number relative to flow behind the moving shock wave. |
| $M_s$ | = | Shock wave Mach number at the launch tube exit, assumed parameter. |
| $P$ | = | Pressure, N/m$^2$. |
| $t$ | = | Time, ms. |
| $u$ | = | Velocity, m/s. |
| $r$ | = | Density, kg/m$^3$. |
| $\gamma$ | = | Ratio of specific heats. |

Subscript

| | | |
|---|---|---|
| $P$ | = | Projectile. |
| $S$ | = | Shock wave. |
| $1,2$ | = | Downstream and upstream of the moving shock wave. |

## I. Introduction

The unsteady flow-fields induced by the projectile when it is launched from a barrel attracted attention of many researchers since it involve various complicated fluid dynamic processes associated with its motion[1-9]. The primary flow disturbances which make the flow field quite unsteady and complex are the flow interfaces in the primary blast wave[10-13] and its interactions[14]. On the other hand, there are projectile flow-field interactions which affect the flight stability of the moving projectile in the flow field. Depending on the Mach number of the jet exiting from the launch tube ahead of the projectile, a secondary shock wave may form behind the primary blast wave[14]. The projectile interacts with the secondary shock wave while it is moving towards the primary blast wave resulting in a drastic change in its aerodynamic characteristics. Another major interaction is the projectile primary blast wave interaction when the projectile catches up the primary blast wave and overtakes it. Later the flow field becomes more complicated when the secondary blast wave overtakes the projectile and interacts with the primary blast wave. These consecutive interactions lead to the so called projectile overtaking problem[3] which has a significant influence on the unsteady aerodynamics characteristics of the projectiles in such flow fields. In addition to these, there are various flow-wave interactions in the flow field which may have coupled effects on the aerodynamic characteristics of the projectile.

Jiang et al[1] have performed a computational analysis to study the fluid dynamics of a moving cylindrical projectile in an unsteady inviscid flow field. They noticed that the interaction between primary blast wave and bow shock wave generated due to the projectile movement in the unsteady flow field is strongly dependent on the projectile speed. Though the fluid dynamics of the flow field was fairly explained in their work, the aerodynamics associated with the flying projectile in the near field has not been addressed. Later, Jiang[2] studied the effect of friction between the projectile wall and the launch tube wall on the wave dynamic processes. He concluded from his numerical results that when the pressure behind the projectile is higher, not only the leading shock of the second blast overtakes the projectile, but the gas behind the projectile does so, resulting in complex wave dynamic processes. He also found that there is a significant increase in the acceleration of the projectile due to this high pressure.

Watanabe[3] studied the fluid dynamics associated with the projectile overtaking problem. They used the one-dimensional theory to analyze various overtaking criteria. They argued that the possible overtaking can be either subsonic or supersonic, depending on projectile relative Mach number.

A computational study on the projectile overtaking a blast wave was performed by Rajesh et al[4]. Their results show that the projectile flow field cannot be categorized based on the relative projectile Mach number, as the Mach number of the blast wave is continuously changing. It is also shown that the aerodynamic characteristics of the projectile are hardly affected by the overtaking process for smaller blast wave Mach numbers as the blast wave will become weak by the time it is overtaken by the projectile. They noticed that the projectile drag coefficient is greatly affected by the unsteady flow structures than the overtaking process. There have been studies to understand the flow field characteristics in the near field of a ballistic range by some researchers[5-7]. They have simulated the flow fields and estimated the unsteady drag coefficients for simple conical shaped projectiles through inviscid simulations.

Schmidt and Shear[8] visualized the flow unsteady projectile flow fields using optical methods. Also there have been analytical studies[9] to predict the flow behavior and the aerodynamics in such flow fields.

Some researchers studied the interaction of various shaped solid objects with the shock waves in steady flow fields[15-17]. Bryson and Gross[15] investigated the interaction of the strong shocks by solid objects of various shapes like cones, cylinders and the spheres in steady flow fields. They found that the experimental results have good agreement with the diffraction theory proposed by whitham[16]. Their experiments showed that the cylinders and the spheres qualitatively have the same kind of interaction pattern.

The computations carried out so far have mainly been inviscid which could hardly capture the actual fluid dynamics of such flow fields. For example, the diffusion of slip lines present in the flow field due to the viscous effect, which may have a deterministic effect on the flow field and projectile aerodynamic characteristics[14]. Moreover, the shock wave-boundary layer interaction is of great importance in such flow fields which may affect the projectile aerodynamic characteristics significantly. Viscous simulations have been carried out by Ahmadikia and Shirani[18] to study the axi-symmetric projectile overtaking problem at transonic and supersonic conditions. Their results show that during the time the projectile overtakes the moving shock wave, the drag force decreases and even becomes negative possibly due to the pressure difference across the shock wave. They also observed that the pressure distribution on the projectile changes when the projectile overtakes the moving shock. However, in their work, they have not considered the blast wave attenuation and its effect on the projectile overtaking problem. In the experimental studies conducted by Sun and Takayama[10, 14] on the shock diffraction in the absence of projectile, it is observed that the viscosity does not affect either the shock structure or the threshold value of the initial blast wave Mach number for the shock diffraction to occur. However, it is found that viscosity affects the structure of the slip lines and the flow field near the wall. Even though there were many attempts to investigate the fluid dynamic phenomena in the near field, most the results have been rather isolated and ambiguously conjectured.

This work here is hence aimed at capturing the fluid dynamics and aerodynamics of the unsteady projectile flow field dynamics in the near field of a launch tube. Two projectile configurations have been employed to study the launch dynamics. A moving grid method is employed to computationally simulate the flow fields. The computations have been performed for various projectile and blast wave Mach numbers. Both inviscid and viscous simulations have been carried out to depict the real flow field features in the vicinity of the launch tube.

## II. Computational methodology

The computational study has been performed using a commercial software CFD-FASTRAN, which makes use of density-based finite volume method that solves the two-dimensional axi-symmetric Euler equations for inviscid simulation[19]. For viscous simulations, the Navier-Stokes equations with appropriate turbulence modeling are solved. Standard *k-E*psilon model is used to model the turbulence. It is a two equation model based on Launder and Spalding (1974) to model the turbulence[20]. This equations model the turbulent kinetic energy and turbulent dissipation rate[19, 21]. The domain is discretized using structural chimera mesh system. For simulating the projectile motion, the chimera mesh scheme allows the overlapping of one zone over the other. The communication between the chimera cells and the overlapping cells is established through a tri-linear interpolation[19]. The projectile is identified as the moving body with six degrees of freedom. The projectile motion is modeled with Euler's equations of motion which is numerically solved at every time step and it requires the physical information of the projectile such as mass and moment of inertia. The forces and the moments are supplied by the flow solver for the motion model. The position, velocity and acceleration at the next time instants are estimated using six degrees of freedom algorithm based on Euler's equation of motion. The projectile has length of 50 mm, diameter of 20 mm, mass of 50 grams and the half-cone angle for the conical projectile is 30°. Since CFD-FASTRAN simulates any axisymmetric problem as a 4° sector of the entire domain, and the mass of the projectile sector is 5.55 E-4 kg and the three principal components of moment of inertia are $I_{xx}$=2.78E-8 kg-m$^2$, $I_{yy}$=$I_{zz}$=1.296E-7 kg-m$^2$. The gas used in the simulations is air with value of specific heat ratio of 1.4.

### A. Computational domain and boundary conditions

The computational domain, the boundary condition and the configuration of the projectiles for the present study are illustrated in Fig.1. The computational domain and the conditions that are used here are the same as that used by Rajesh et al[4].

The solver uses Van Leer's flux vector splitting scheme for spatial discretization which is of first order accuracy. The spatial accuracy of the scheme is increased to higher orders with the use of Osher-Chakravarthy flux limiter[22]. The time integration is carried out using point Jacobi fully implicit scheme.

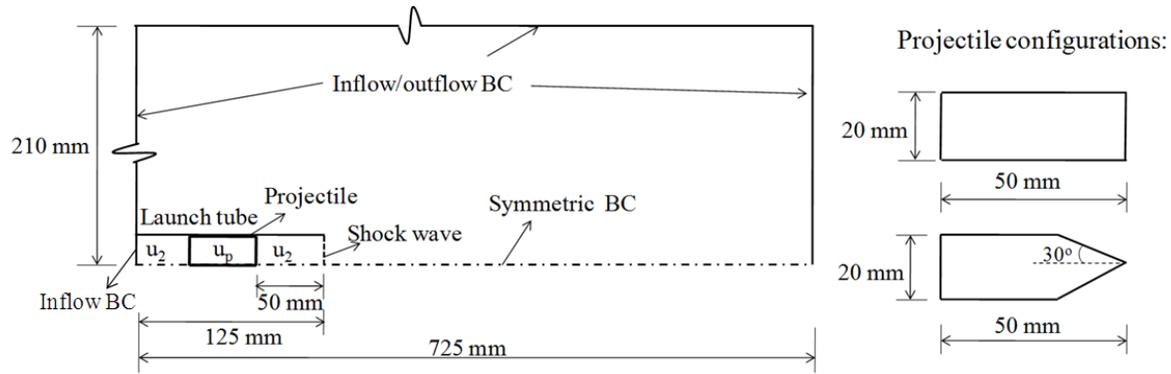

**Fig 1: Computational domain, boundary conditions and Projectile configurations.**

The inflow/outflow boundary conditions with pressure and temperature values of standard atmosphere used at the far fields are suitable for representing far-field boundary condition for external flow problems. For supersonic outflows, the values of the variables at boundary are extrapolated from the interior and the boundary does not affect the upstream flow conditions. The far-fields are given at a distances of 210 mm from the projectile. This distance has been finalized after having observed that no strong wave reaches this boundary in the near-field, in many initial simulations. Initially the moving shock wave at the exit of the launch tube is assumed to have a Mach number $M_s$. This Mach number is one of the parameters of the study and so the computational model does not include the effects of expansion waves moving into the launch tube, after projectile exit, and the resulting modulation of launch tube blow down; rather the outflow from the launch is an assumed parameter.

The projectile is kept inside at a distance of 50 mm behind the shock wave. The velocity of the projectile and the flows ahead and behind the projectile are assumed to be in the same direction and magnitude of that of the downstream flow behind the moving shock wave which is at the exit of the launch tube. This assumption leads to the condition that the projectile base pressure during its travel in the launch tube is constant. However, in a real gas gun with a chambrage, an unsteady shockwave due to the diaphragm rupture is moving into the launch tube and gets reflected on the projectile base. Moreover, an expansion can also be initiated due to the sudden acceleration of the projectile. The projectile experiences drastic fluctuations in the acceleration due to multiple reflections owing to these wave excursions. However, it has previously been observed[23] that the fluctuations in the projectile base pressure damp out very fast and the velocity doesn't seem to be affected greatly[5]. This is dependent on many factors such as the diaphragm rupture pressure, area ratio of the pump tube to launch tube, projectile mass and diameter, and the type of the gas, etc[23, 24]. The projectile mass used here is 50g and hence the possibility of drastic fluctuations in the acceleration due to the wave reflections on its base may be considerably reduced due to the inertia effects and may not affect the fluid dynamic phenomena in the near field.

Based on the above arguments, it is reasonable to assume that the base pressure on the projectile is constant, and temperature and pressure values of the inlet boundary condition of the launch tube are same as that of the flow conditions behind the projectile. Overset boundary conditions are assigned to the interface boundaries between chimera mesh and the background mesh. Adiabatic wall boundary conditions are given to the launch tube walls and the walls of the projectile.

## B. Grid and time independence study

In the case of inviscid simulation, the same grid system and time step as that of Rajesh et al[4] were employed. Fig.2a shows the validation of the numerical scheme used by Rajesh et al with a particular case from Jiang etal[1] of which the shock wave at the launch tube exit ($Ms$) is 3.0. The comparison of the results shows that all the fluid dynamic phenomena are well captured by the numerical scheme employed.

In order to determine the optimum mesh size and the time step size for viscous cases, the time and grid independence studies have been performed as shown in Figs. 2b and 2c. For the viscous analyses, the grid independence study has been performed with the number of cells of 460,000, 300,000 and 150,000 for the case of $M_{p1}$=1.25. The time independence study has been performed with time steps of 2E-9, 3.5E-9 and 6.5E-9 for the case of $M_{p1}$=1.25.

The acceleration history of the projectile is shown for various time steps in Fig.2a and for various numbers of cells in Fig. 2b. There are hardly any variations in the plots. Hence, for lower blast wave Mach numbers, the time step chosen was 3.5E-9 second and for higher Mach numbers the chosen time step was 2E-9. This is to assure the stability of the flow solver which is found to become degrade when higher Mach numbers are chosen. Similarly, the number of cells chosen for all the Mach numbers is 460,000.

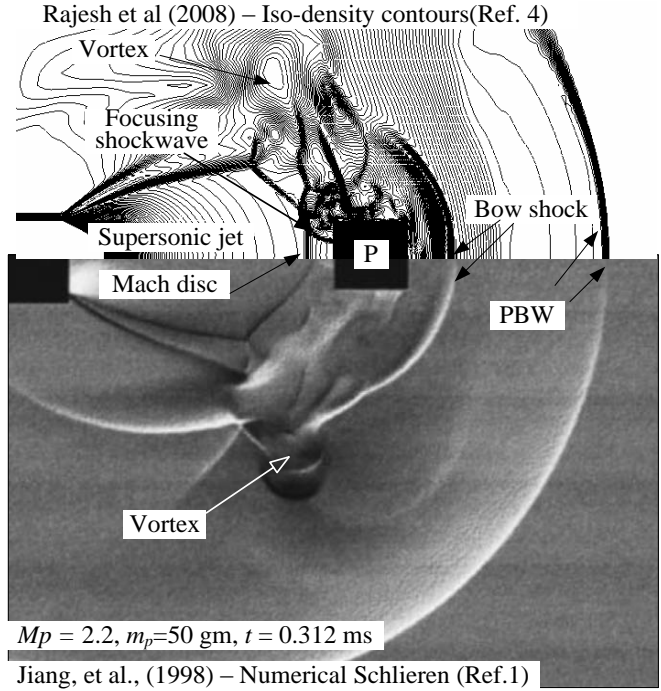

**Fig. 2.a: Comparison of iso-density contours with numerical schlieren.**

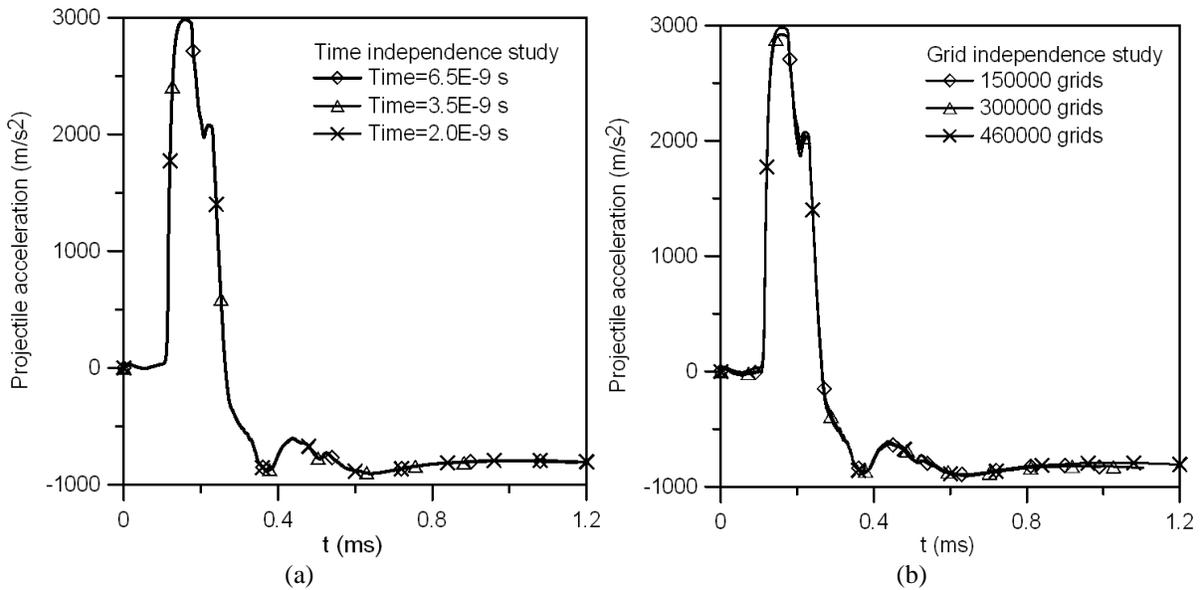

(a) (b)
**Fig. 2.b: Time independence study                c: Grid independence study.**

## III. Results and discussions

### A. The projectile overtaking phenomenon-revisited[4]

There are two distinct flow regimes in the launch dynamics based on the overtaking phenomenon; one is the possible overtaking condition and the other being impossible overtaking condition. The possible overtaking is further classified in to supersonic overtaking and subsonic overtaking condition. When relative Mach number $M_{p2}$ is greater than 1.0, the projectile is moving supersonically relative to the flow behind the precursor shock wave with its Mach number $M_s$. Meanwhile, when $M_{p2}$ is less than 1.0, the projectile is moving subsonic with respect to the flow behind the blast wave. The conditions in which $M_{p2}$ is greater than 1.0 or less than 1.0 are referred to as supersonic and subsonic overtaking conditions, respectively. If the projectile Mach number is less than the blast-wave Mach number ($M_{p1}<M_s$), the projectile cannot overtake the blast wave. This is known as the impossible overtaking condition[4].

The main characteristic of the supersonic overtaking condition is that the disturbance due to the projectile motion cannot travel upstream and affect the blast wave. Moreover, in supersonic overtaking condition, there will be a standing shock wave in front of the projectile. However, in case of subsonic overtaking condition, the projectile travels at subsonic velocity behind the blast wave and the disturbance wave due to the projectile motion may affect the blast wave ahead. In real situations, the blast wave attenuates faster and the projectile relative velocity increases to supersonic condition and the subsonic overtaking condition may not happen in most of the cases.

### B. Projectile aerodynamic characteristics - inviscid flow

The acceleration histories of the cylindrical and the conical projectiles for both the inviscid and viscous cases are analyzed for various Mach numbers ranging from $M_s$=1.4 to $M_s$=3.0. Figure 3a shows the typical acceleration histories of the projectiles for the Mach numbers $M_s$=2.0. The acceleration history of the cylindrical projectile passing through unsteady flow structures for various Mach numbers is discussed in detail for the inviscid case in Rajesh et al[4]. Various state points are marked in the Fig. 3 based on the projectile-flow field interactions to bring out the salient features of the aerodynamics characteristics of the projectile. In order to investigate configuration effects on the projectile aerodynamic characteristics, the drag coefficient histories of the cylindrical and conical projectiles are computed and compared, where the drag coefficient ($C_d$) is defined[4] as

$$C_d = \frac{2D}{\gamma p_1 M_{p1}^2 A_p}$$

Where $D$ is the drag force, $\gamma$ is the ratio of specific heats, $P_1$ is the ambient pressure, $M_{p1}$ is the projectile Mach number relative to ambient conditions, $A_p$ is the projected frontal area of the projectile.

**1.** *Projectile-secondary shock interaction*

Before the projectile exits from the launch tube, the secondary shock wave forms with in the primary blast wave as a result of the shock wave diffraction at the convex corner of the launch tube and the jet Mach number is greater than the threshold Mach number[14] for the generation of the secondary shockwave. This is followed by an under-expanded jet. Once the head of the projectile comes out of the launch tube, the front face of the projectile starts interacting with the low pressure, under expanded jet resulting in a pressure difference between the front and rear side of the projectile. This causes an increase in the acceleration of the projectile from state "a" onwards, as shown in Fig. 3a. The acceleration of the projectile continues to increase till it interacts with the secondary shock wave which is marked as state "b". The corresponding state points are marked in the drag coefficient history as shown in the Fig. 3b. The acceleration of the projectile reaches a maximum at this state "b" and the corresponding drag coefficient reaches minimum as shown in the figure.

*a. Configuration effects*

Until state "b", the drag coefficient histories of the cylindrical and conical projectiles are qualitatively and quantitatively similar as shown in Fig. 3. This is because the shapes of the rear ends of both the projectiles are the same and the aerodynamic characteristics till this point are determined mainly by the jet behind the projectile. At state "b", there is a sudden drop in the acceleration. This is the point where the projectile interacts with the secondary shock wave and enters a flow field, where the projectile Mach number relative to flow behind the moving shock wave ($M_{p2}$) becomes supersonic. This sequence of events can be seen in the Mach number contours of the Figs. 4a, 4b and 4c. From the state "c" to "d", there is a fluctuation in the acceleration of the cylindrical projectile. This can be attributed to the formation of the bow shock wave in front of the cylindrical projectile[4]. Similar trend is also observed in the drag coefficient of the cylindrical projectile.

However, for the conical projectile this fluctuation in the drag coefficient corresponding to the interaction with the secondary shock wave from state point "c" to "d" is not observed as seen from Figs. 3a and 3b. This shows that the projectile interaction with the secondary shock in the case of the cylindrical projectile is quite different from that of the conical projectile. This is due to the aerodynamic shape of the conical projectile and its effect on the flow field. This effect can be clearly seen in Figs. 4b and 4c where the contours of the projectile Mach numbers are shown. Similar trends as of those explained above are observed in the acceleration and $C_d$ curves of $M_s=2.5$ and are shown in Figs. 5a and 5b. The corresponding Mach contours are shown in Fig. 6.

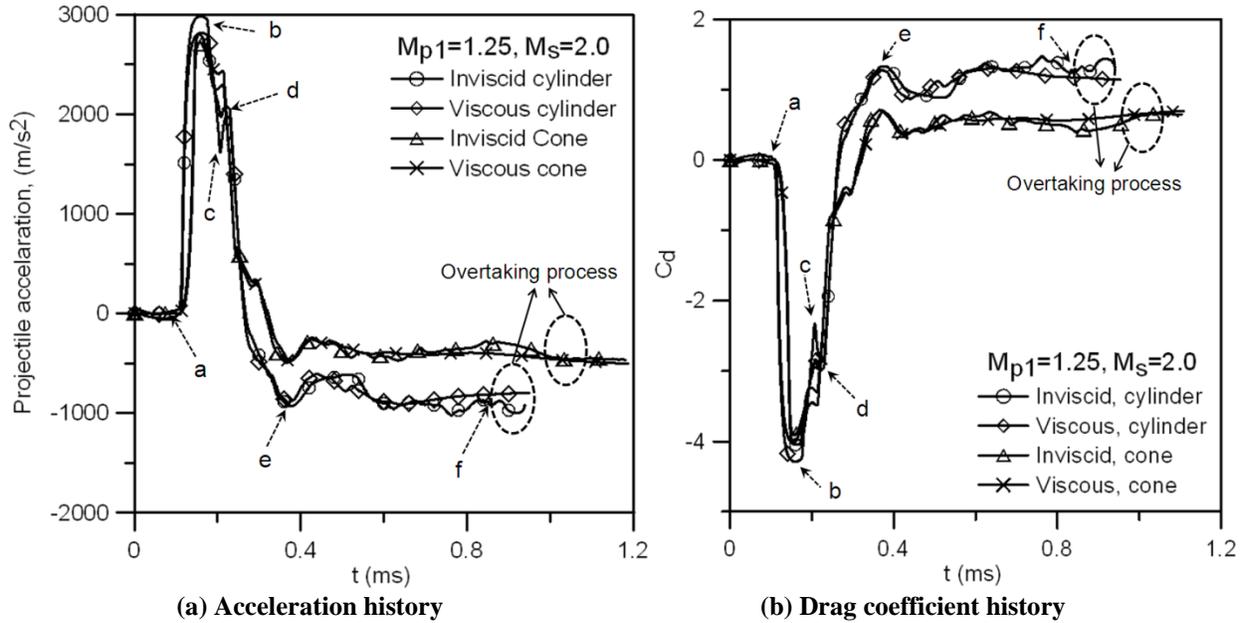

(a) Acceleration history  (b) Drag coefficient history

**Fig. 3: Acceleration and drag coefficient histories of the cylindrical and conical projectiles for inviscid and viscous simulations for Mach number $M_s=2.0$ and $M_{p1}=1.25$.**

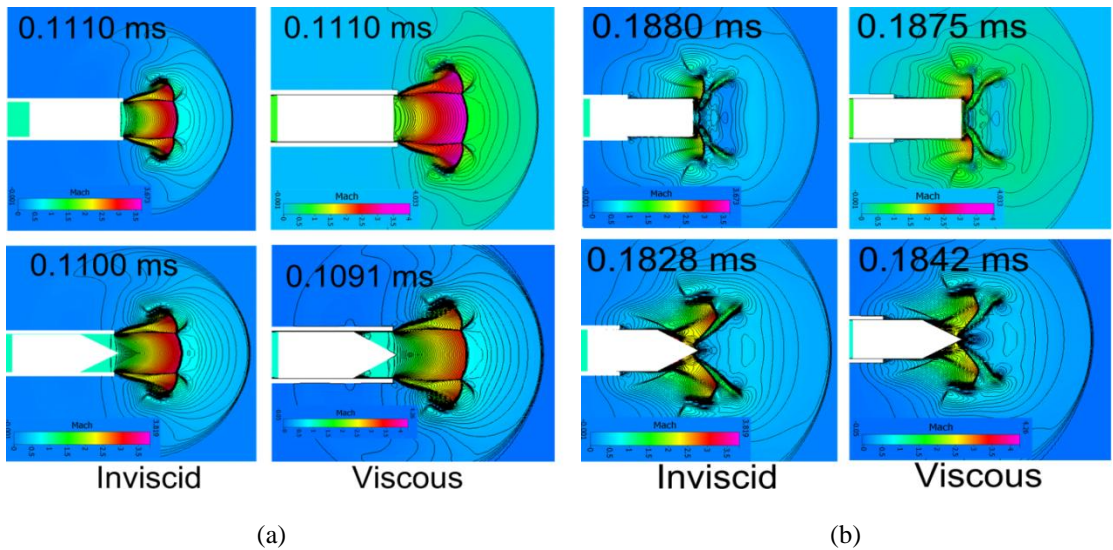

(a)                (b)

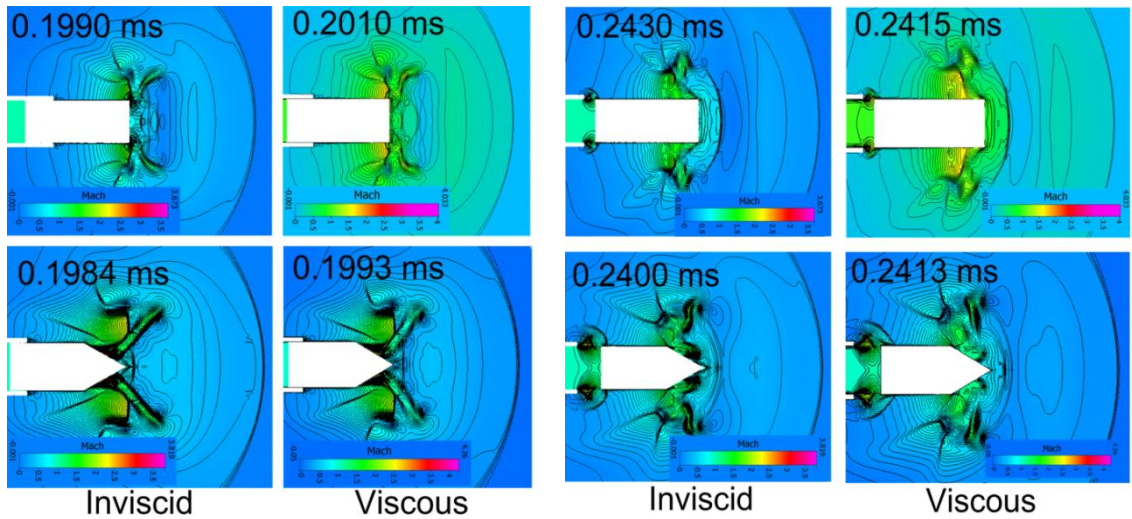

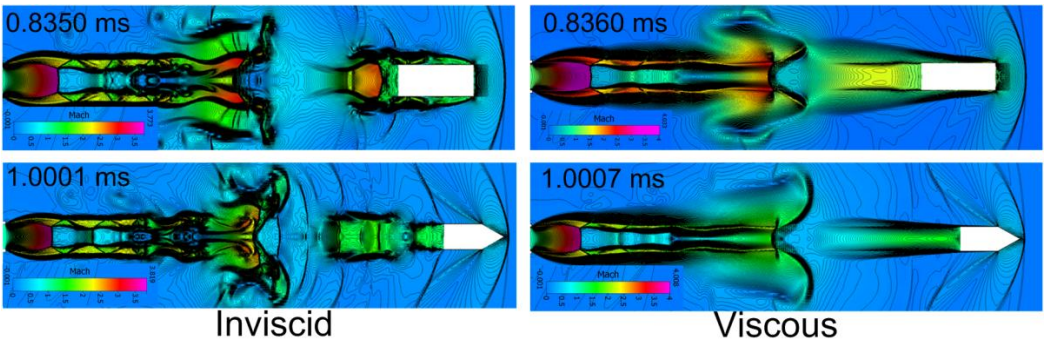

Fig. 4: Mach contours for $M_{PI}$=1.25 and $M_s$=2. Mach number scale range: 0-4. The axial and the radial dimensions of the computational space are 725 mm and 210 mm.

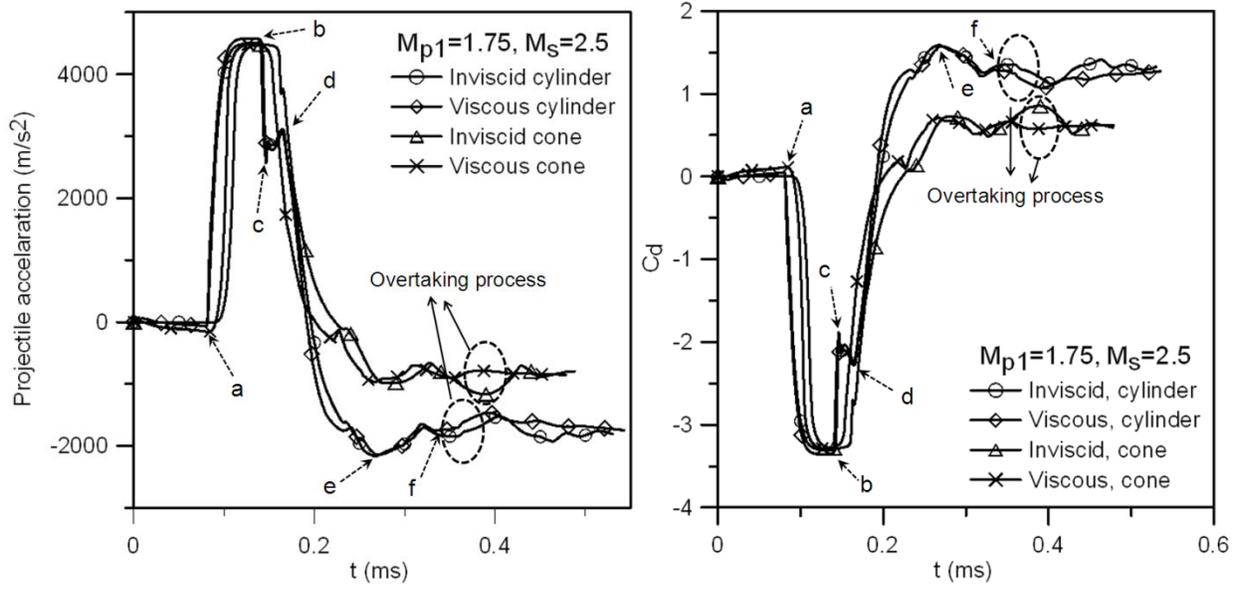

**(a) Acceleration history**  **(b) Drag coefficient history**

**Fig. 5: Acceleration and drag coefficient histories of the cylindrical and conical projectiles for inviscid and viscous simulations for Mach number $M_s$=2.5 and $M_{p1}$=1.75.**

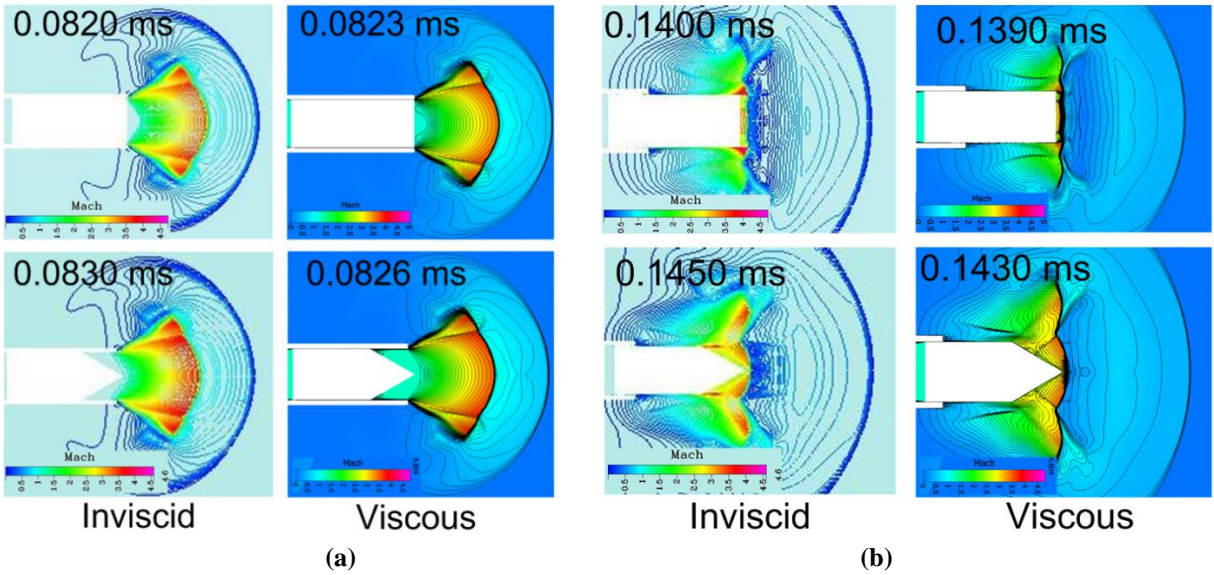

Inviscid  Viscous  Inviscid  Viscous
(a)  (b)

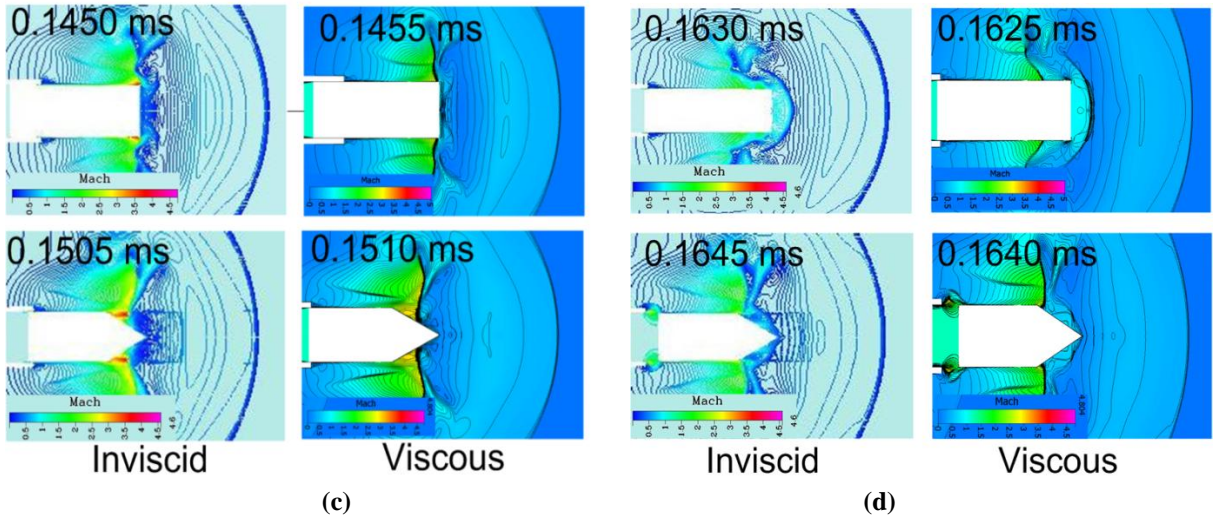

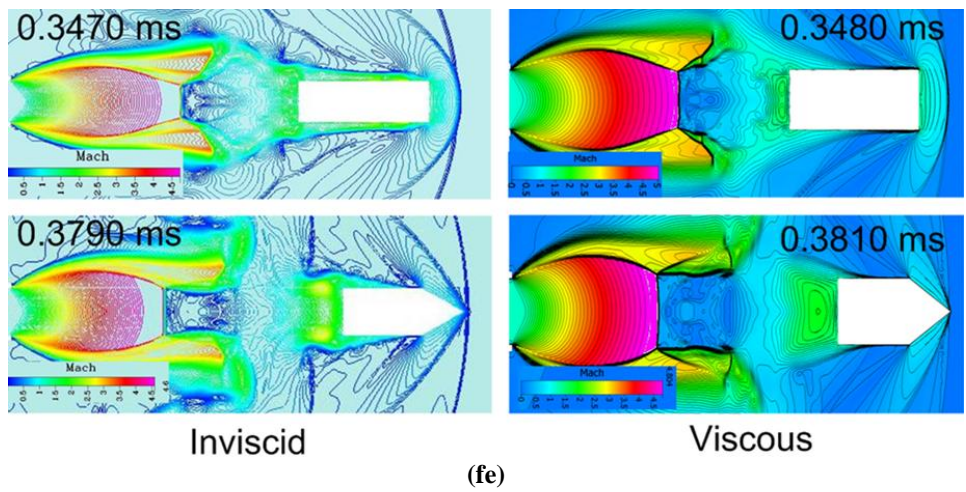

**Fig. 6:** Mach contours for $M_{P1}$=1.75 and $M_s$=2.5. Mach number scale range: 0-5. The axial and the radial dimensions of the computational space are 725 mm and 210 mm.

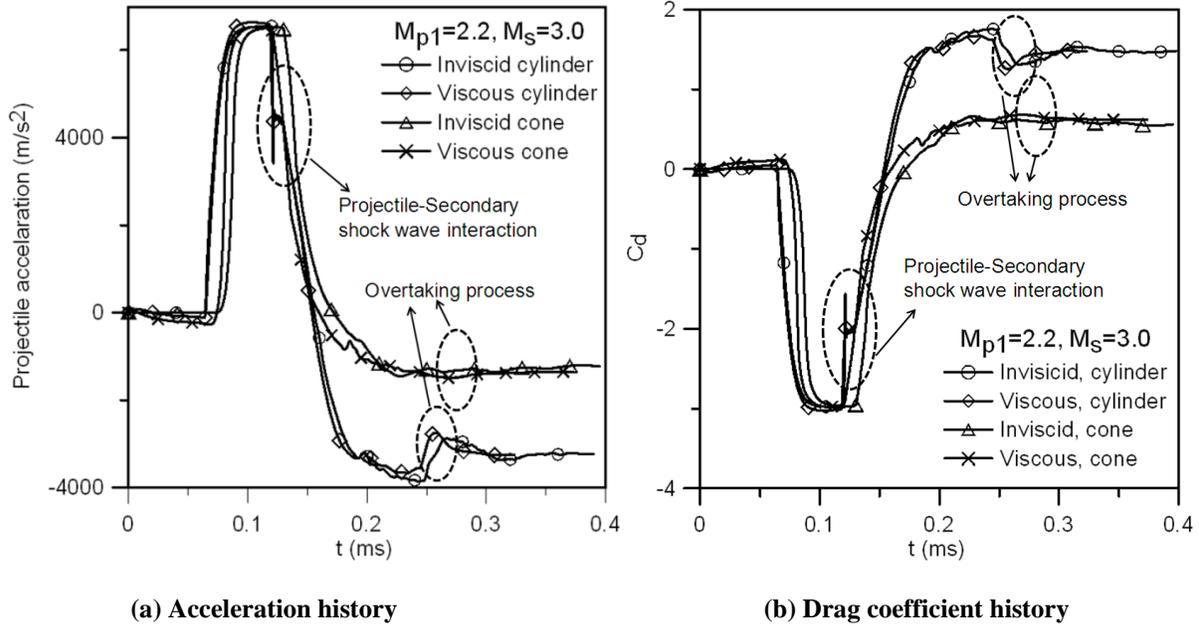

**(a) Acceleration history**        **(b) Drag coefficient history**

**Fig. 7: Acceleration and drag coefficient histories of the cylindrical and conical projectiles of inviscid and viscous simulations for Mach numbers $M_s$=3.0 and $M_{p1}$=2.2.**

The smooth interaction between the conical projectile and the secondary shock wave reveals the absence of excursion of waves when the projectile is heading towards the bow shock wave, as in the case of a cylindrical projectile[7]. From the state "e" onwards, the projectiles undergo a steady deceleration. This is due to the fact that the driving force from the jet in the secondary blast wave diminishes as the projectile moved sufficiently far from the launch tube exit. It can also be observed that the cylindrical projectile decelerates faster than the conical projectile. This verifies that the strength of the bow shock wave being developed in front of the cylindrical projectile is higher that of the conical projectile.

After carefully examining the projectile-wave interaction within the blast wave, it is noted that a more detailed analysis of this interaction and its effect on the projectile overtaking will be worthwhile. The interaction of projectile with the secondary shock wave is the point where the classification of the flow fields based on the relative projectile Mach number becomes important. When the projectile passes through the secondary shock wave, the relative projectile Mach number increases to the supersonic state. It is noticed that during the projectile-secondary shock wave interaction, the dynamics of the shock wave is highly dependent on the configuration of the projectile. This can be clearly seen from the Figs. 4b and 6b. For the case of cylindrical projectile the shock wave preserves its shape until the interaction. But for the conical projectile, the characteristics of the secondary shock wave change as it is approached by the conical face of the projectile. This is mainly due to the turning of the flow field in the vicinity of the conical face of the projectile. This causes the normal shock wave to evolve as an attached oblique shock wave in order to meet the downstream flow conditions, provided the relative Mach number is large enough to produce an attached oblique shock wave for the half angle of the cone.

### *b. Mach number effects.*

The Mach number decides the nature of the interaction between the projectile and the secondary shock wave. It is observed for the cylindrical projectile that the higher the Mach number, the stronger the variation in the drag coefficient of the projectile. This can be clearly seen through Figs. 3b, 5b, 7b that the variation in $C_d$ is more in the case of $M_s=3.0$ compared to that of $M_s=2.5$ and $M_s=2.0$ However, for the case of conical projectile, the increase in the Mach number does not seem to affect the drag coefficient much during the interaction as seen in the figure 7b.

### *2. Projectile overtaking process.*

When the cylindrical projectile is overtaking the blast wave, the bow shock wave in front of the projectile interacts with the blast wave and forms the familiar triple point on either side of the projectile. The overtaking phenomenon starts at state "f" where the bow shock wave in front of the projectile just starts interacting with the blast wave which is marked in Figs. 3a and 3b. This is followed by a distinct triple point formation where the fluctuations in the projectile drag coefficient are very small. This is probably due to a weakened blast wave owing to the attenuation of the blast wave to an extent where the pressure jump across the wave is quite smaller than that needed to create a drastic fluctuation in the drag characteristics of the projectile. This is in contrast with Watanabe's[3] results where the blast wave maintains a constant strength throughout the overtaking process. In real situations as described here, the blast wave strength is diminishing rapidly and hence it alters the flow conditions behind the blast wave. This leads the only possible overtaking condition to be supersonic as the relative projectile Mach number has reached a value greater than 1 and is still increasing. It is also observed in the present simulations that the overtaking phenomenon does not seem to greatly affect the aerodynamic characteristics of the projectile for the Mach numbers under consideration.

*a. Mach number and configuration effects.*

In Figs. 3b, 5b corresponds to $M_s$=2.0 and $M_s$=2.5, it is seen that there are no significant variations in the drag coefficients of both the projectiles during the overtaking process. However, in Fig. 7b, where the projectile drag coefficient history is plotted for $M_s$=3.0, there is a significant variation seen during the overtaking process in the case of the cylindrical projectile. This is due to the fact that for higher Mach numbers, the blast wave attenuation does not significantly weaken the pressure jump across the blast wave at least in the near field. It can then be inferred that, for higher Mach numbers, the overtaking process affects the projectile aerodynamic characteristics greatly compared to the unsteady flow structures behind the blast wave. It is also seen that this fluctuation in the drag coefficient at the time of overtaking is not observed in the case of the conical projectile, compared to the case of cylindrical projectile. The variation in the drag coefficient in the case of the cylindrical projectile can be attributed to the excursion of the compression waves between the bow shock and the front surface of the cylindrical projectile, which on interaction with the blast wave, strengthen the latter and creates a large pressure jump. On the other hand, this excursion mechanism is absent in the case of the conical projectile due to its aerodynamic shape and there will be a smooth interaction with the blast wave and hence the projectile drag coefficient will not be affected greatly.

*3. Projectile overtaking criteria*

*a. x-t-diagrams:*

To identify the projectile blast wave motions and the overtaking process, the $x - t$ diagrams are plotted and shown in Fig. 8a, and 8b. The x-t diagrams of the projectile are constructed plotting the path of the tip of the projectiles with time. For the cylindrical projectile, the tip of the projectile never meets with the path of the blast wave as there is shock stand-off distance. This is true for a detached shock on the conical projectile too. The overtaking point in this case is defined as the merging of the bow shock/detached shock with the precursor shockwave. On the other hand, the overtaking point of a conical projectile is the point at which the tip of the projectile and the attached shockwave meets with the precursor shockwave. The numerical values for overtaking time shown in the Fig. 8 are based on these concepts. The blast wave starts from the launch tube exit and the projectile tip starts at 50 mm behind the shock wave. As the time progresses, the paths of the projectile and blast wave are converging and the point where the tip of the cylindrical projectile becomes closest to the path of the blast wave is identified as the overtaking point and is shown in the Figs. 8a, 8b. This is due to the shock stand-off distance in front of cylindrical projectile.

In the case of the cone, the shock wave formed in front of the projectile can be attached or detached depending on the critical Mach number $M_{cr}$ for the projectile half cone angle $\theta_{max}$. If the relative projectile Mach number is less than the critical Mach number $M_{cr}$, the shock wave is detached. This is particularly true for $M_s$=2.0 where $M_{p1}$=1.25. It can be seen that the shock wave formed in front of the conical projectile is a detached one and the shock stand-off distance is identified from the *x-t* diagram. However, the shock stand-off distance for the cone is much smaller compared to that of the cylindrical projectile. For the attached shockwaves, the overtaking point is the point at which the tip of the cone meets with the precursor shockwave path. The overtaking points given in terms of milliseconds in the plots are based on the above concepts. From the Figs. 8a and 8b, it is clear that the configuration does not affect the overtaking path except that the bow shock wave standoff distance for the cylinder is much larger than that of the cone.

### *b. Mach number and configuration effects*

To get the possible overtaking regimes of the projectile at various Mach numbers, the variations of $M_s$ and $M_{p2}$ are plotted against time in Figs. 9a-b. It clarifies the effect of the attenuating blast wave on the overtaking process. It can be seen that during the whole overtaking process, the blast wave Mach number is varying from an impossible overtaking ($M_{p1}<M_s$) condition to possible one ($M_{p1}>M_s$)[4]. The overtaking is impossible until time $t_1$ for the case of $M_s$=3 and time $t_2$ for the case of $M_s$=2.5 and time $t_3$ for the case of $M_s$=2.0, since the blast wave travels faster than the projectile in this regime, and after this time the overtaking becomes possible due to the blast wave attenuation and consequent increase in $M_{p2}$[4]. Overtaking is never possible for the case of $M_s$=1.4, $M_{p1}$=0.57 as the projectile Mach number is always subsonic. The blast wave attenuations for conical and cylindrical projectile are fundamentally of similar nature. The projectile Mach number does not change significantly in the near field. This confirms that the configuration changes do not affect the overtaking regimes.

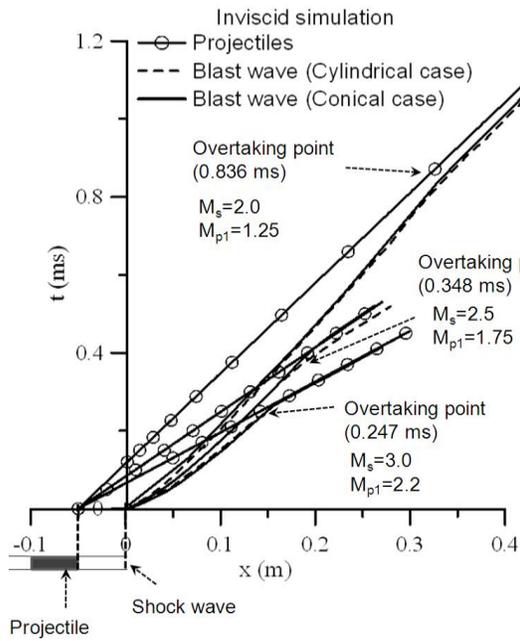 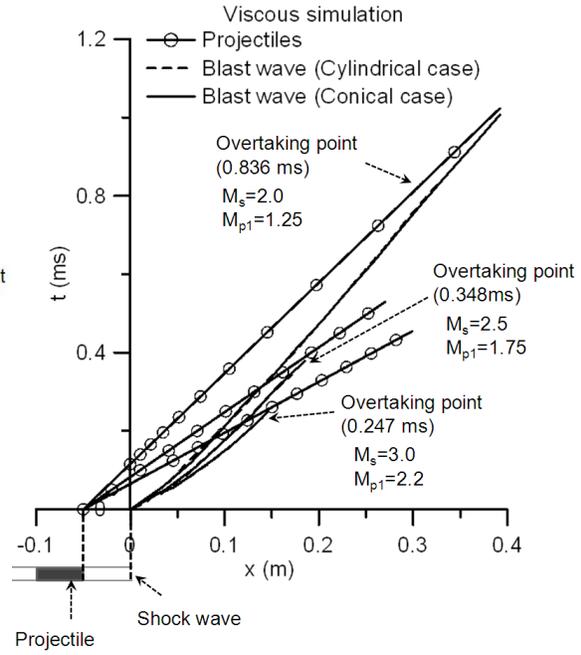

**(a)** Inviscid simulation, configuration effects  **(b)** Viscous simulation, configuration effects

**Fig8:** *x-t* diagrams of the projectile and blast wave

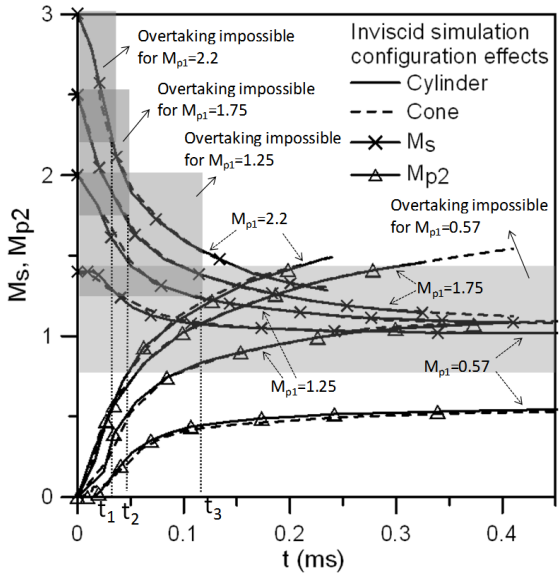 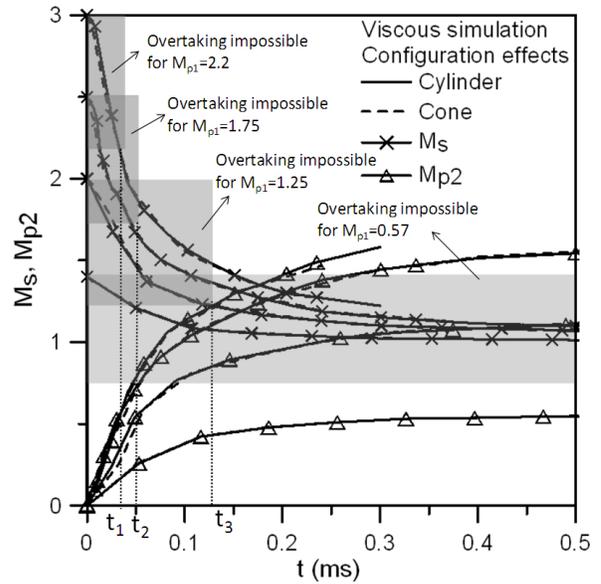

**(a)** Inviscid simulation, configuration effects  **(b)** Viscous simulation, configuration effects

**Fig. 9:** variation of $M_s$ and $M_{p2}$ with respect to time

**C. Viscous effects**

The effects of viscosity on the flow field as well as on the projectile aerodynamics characteristics are analyzed here. The turbulence is modeled using standard $k$-$\varepsilon$ model along with the viscosity.

*1. Projectile-flow field interaction.*

A predominant feature of the viscous flows in such flow fields is the diffusion of the vorticity as seen through the Fig. 10 a-e. In order to bring out the effect of viscosity on the flow structures clearly, the vorticity contours are shown here. It is known that the viscous effect is a dominating factor in determining the structure of the slip lines in such flow fields[14].

The presence of viscosity enhances the diffusion of vorticity and it alters the shock wave structures. This process is relatively slow and its effect can be seen only at the instant when the projectile overtakes the blast wave, at a sufficiently large distance from the launch tube exit. The diffusion of the vorticity structures during the projectile overtaking phenomenon can be seen clearly in Fig. 10e. During the initial stages, as seen from Figs. 10 a-c, the evolution of the slip lines seems to occur in similar manner for both inviscid and viscous cases. The diffusion is observed in the shear layers originated from the lip of the launch tube and from the corner of the projectile wall in viscous case.

One notable difference between the viscous and the inviscid cases is the attenuation of the shock structures behind the projectile due to the viscous effects. This can be seen in Figs. 4e and 6e where Mach number contours are shown. The shock structures in the jet near the secondary blast wave are significantly affected in the viscous case. It is also noticed that the shock structures seen beneath the shear layers originating from the front corner of the projectile in the inviscid case, are not observed in the viscous case. This confirms the fact that the viscous effects are responsible for the vorticity diffusion which is highly influential in shaping the velocity fields and hence the shock structures. This is clear in Fig. 10e also, where it is seen that the shock structures are significantly altered by the diffusion of vorticity.

*a. Mach number effects.*

Figures 3, 5 and 7 show the aerodynamic characteristics of the projectile of $M_s$=2.0, $M_s$=2.5 and $M_s$=3.0 for the inviscid and viscous cases of both projectile configurations. For higher Mach numbers, for which the projectile relative Mach number is supersonic with respect to the ambient condition, the aerodynamic characteristics of the inviscid and viscous cases are similar. This shows that the viscosity hardly affects the aerodynamic characteristics of the projectile at least for moderate Mach numbers. However, for a lower blast wave Mach number where the projectile relative Mach number is subsonic, there are large variations in the aerodynamic characteristics of the inviscid and viscous cases as will be discussed in the section D.

## *2. Projectile aerodynamic characteristics.*

Regarding the aerodynamic characteristics of the moving projectile, it is observed that the viscosity plays a less significant role in the regime of Mach numbers. The boundary layer forms over the walls of the projectile that are responsible for the shear or viscous drag, are absent in the case of inviscid flow. On the other hand, the pressure drag on the moving projectile is quite different from that on a steady bluff body in a wind tunnel or a projectile moving in a steady flow field. The projectile experiences drastic pressure fluctuations locally when it goes through the intricate flow interfaces. Hence the pressure drag is the dominant one here compared to the viscous drag in determining the unsteady drag coefficient. This is clearly seen in projectile drag coefficient histories from the Figs. 3b, 5b, and 7b.

It can also be seen in Figs. 9a, 9b that the configuration changes do not affect the possible overtaking regimes for the viscous case as the viscosity does not affect either the blast wave attenuation or the projectile Mach number in the near field to a significant extent.

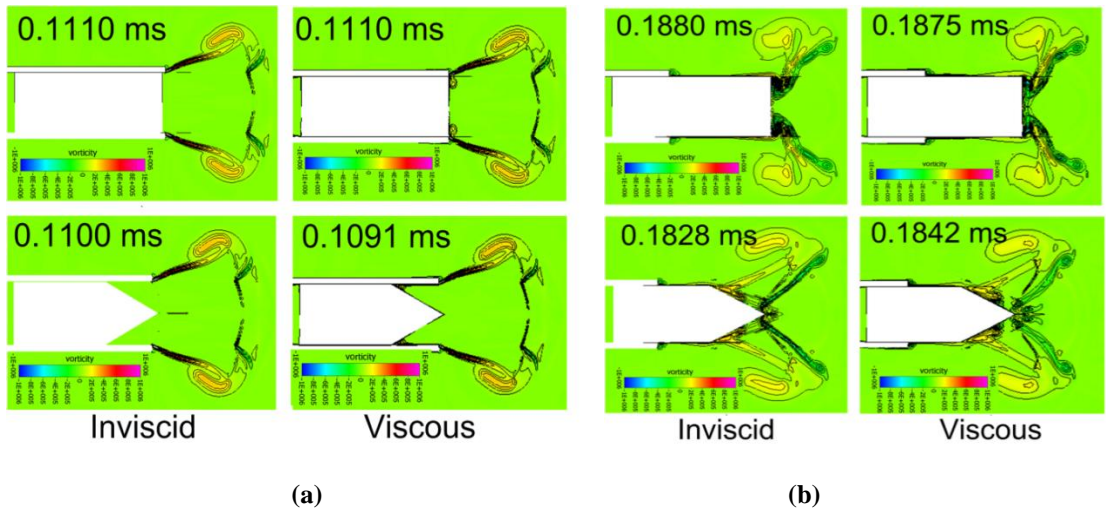

(a)  (b)

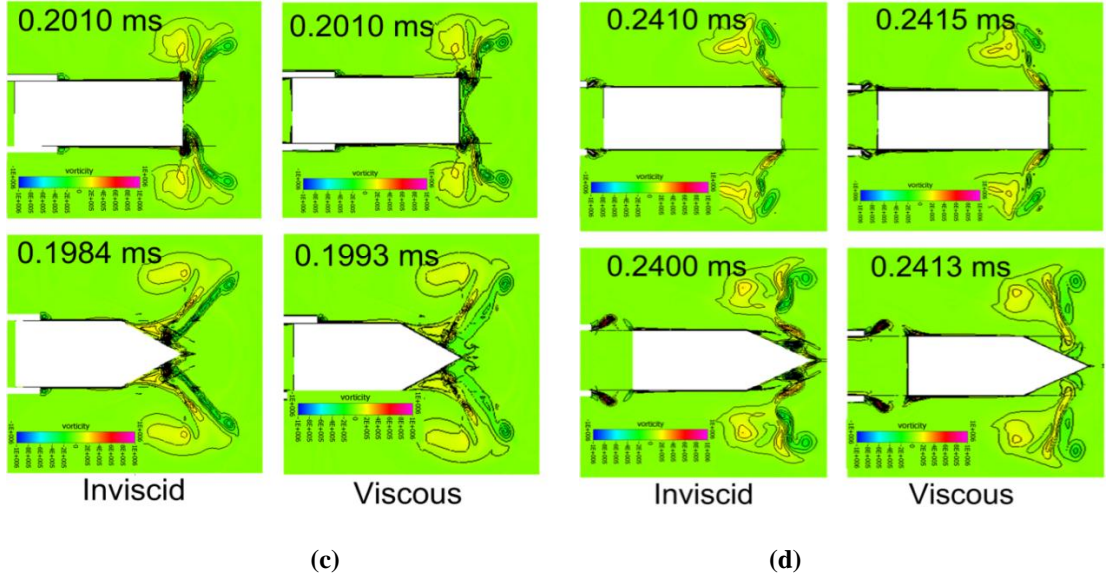

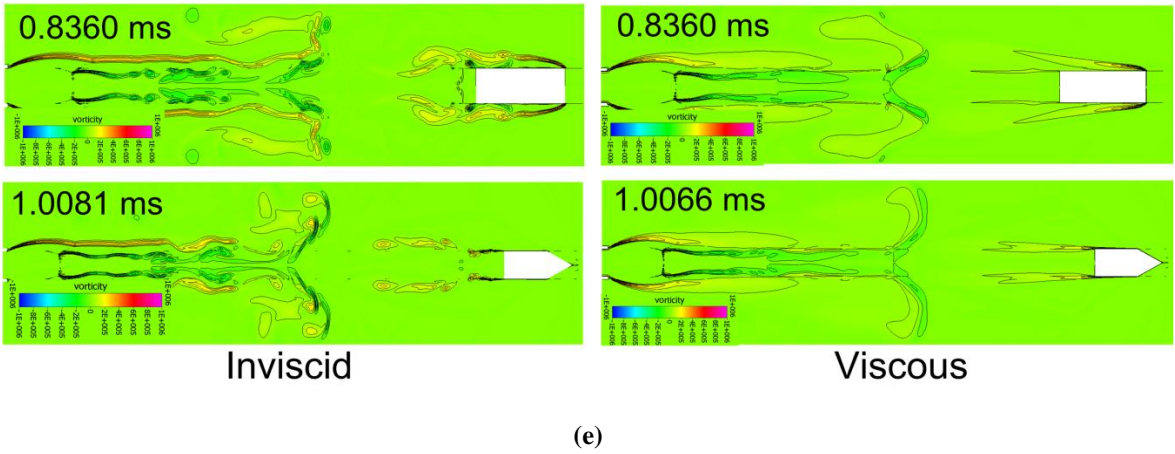

**Fig. 10**: Vorticity contours for $M_{P1}=1.25$ and $M_s=2$. The vorticity scale range: $-10^6$ to $10^6$ s$^{-1}$. The axial and the radial dimensions of the computational space are 725 mm and 210 mm.

### D. The impossible overtaking regime ($M_{p1}=0.57$ and $M_s=1.4$).

In the case of $M_s=1.4$, the projectile Mach number is subsonic with respect to the ambient conditions and the trends in the acceleration and the drag coefficient histories are entirely different from that of the higher Mach numbers, as shown in Figs. 11a and 11b. There are large variations in the acceleration and the drag coefficient histories of the projectiles. These variations are mainly due to the transient flow structures in the flow field.

When the projectile travels inside the launch tube, the shock wave diffraction is followed by the discharge of a subsonic jet. This creates expansion waves at the exit of the launch tube before the projectile discharge as the subsonic jet exits to a low pressure region generated due to the shock wave diffraction. The jet is subsonic and hence the expansion wave can travel upstream into the launch tube towards the projectile face and creates a low pressure in front of the projectile. This pressure difference causes the acceleration of the projectile inside the launch tube itself and this is marked as state "1" in Fig. 11 and the corresponding pressure contour is shown in Fig. 12a. Moreover, the shock wave diffraction results in the formation of moving vortical structures with which the projectile interacts later.

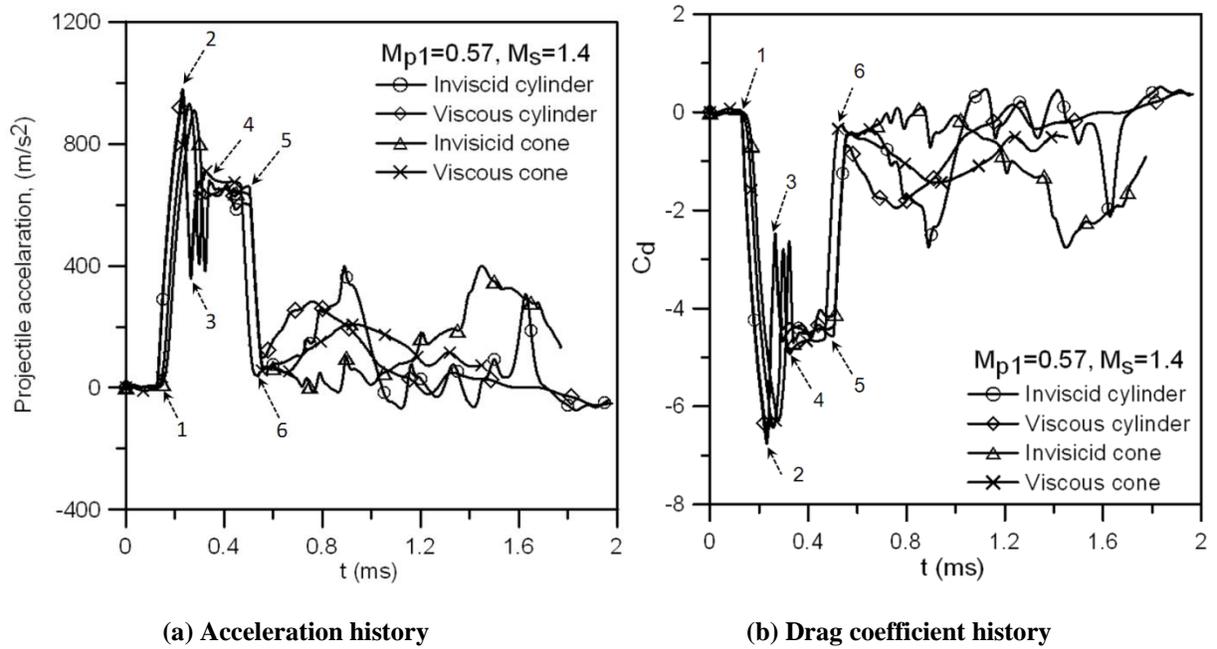

(a) Acceleration history  (b) Drag coefficient history

**Fig. 11:** Acceleration and drag coefficient histories of the cylindrical and conical projectiles for inviscid and viscous simulations for Mach numbers $M_s$=1.4 and $M_{p1}$=0.57.

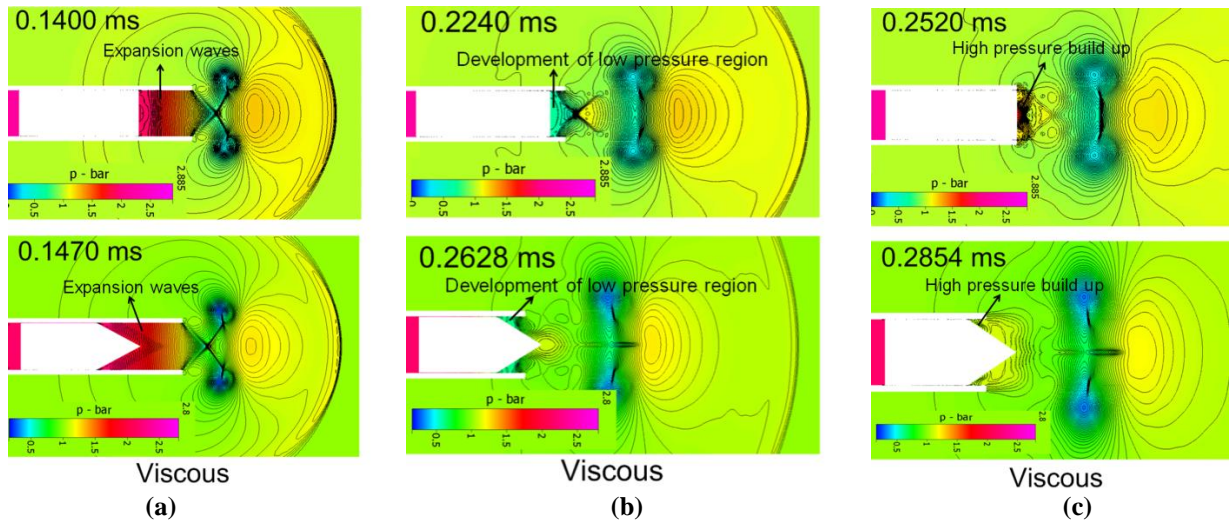

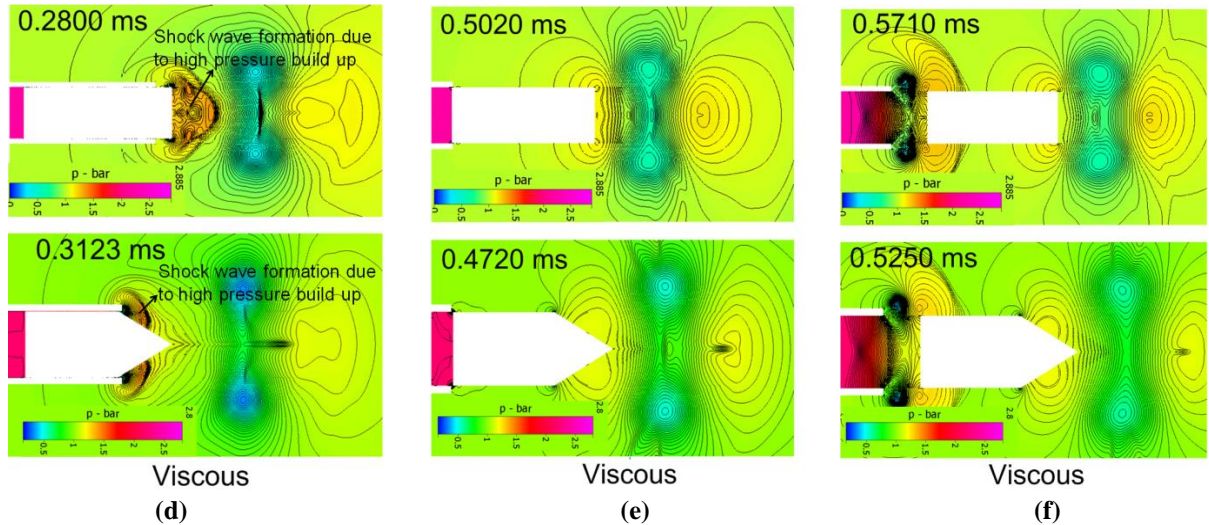

**Fig. 12: Pressure contours of viscous simulation obtained for $M_{P1}$=0.57 and $M_s$=1.4. The pressure scale range: 0.2 to 2.88. The axial and the radial dimensions of the computational space are 725 mm and 210 mm.**

There is a significant increase in the acceleration of the projectile until its head starts coming out of the launch tube. When the projectile head reaches the exit of the launch tube, a further low pressure is created locally near the front face of the projectile just before it interacts with the expansion wave ahead of it. This low pressure region is indicated in Fig. 12b. Following the low pressure formation, a high pressure build up takes place when the projectile encounters pressure difference across the expansion wave. The duration of this process is so short that there is a drastic decrease in the acceleration of the projectile. This process occurs between state "2" to state "3" as shown in Fig. 11. Corresponding pressure contours are seen in Figs. 12b and 12c. The increase in the pressure in front of the projectile can be seen in the Fig. 12c. This high pressure then creates a weak bow shock in front of the projectile as indicated in Fig.12d. The shock wave travels along with the projectile and attenuates resulting in reduction of the pressure drag on the projectile as indicated as state "4" in Fig. 11. During this process the acceleration of the projectile increased from state "3" to state "4" as shown in Fig. 11. From the state "4" onwards there is uniform pressure in front of the projectile leading to a constant acceleration till the projectile completely exits from the launch tube and this process is marked from state "4" to state "5".

Once the projectile exits from the launch tube, the secondary blast wave forms behind the projectile and it sweeps over the surface of the projectile followed by the formation of low pressure region behind the rear face of the projectile. This causes a decrease in the projectile acceleration instantly from state "5" to state "6". Till state "6", the acceleration histories of the projectile show the same trend for both the viscous and inviscid simulation as seen from Fig. 11. The drag coefficient of the projectile shows similar negative trend as that of the acceleration.

From the state "6" onwards which corresponds to the point where the projectile just starts interacting with the moving vortical structures, there are complex interactions among the projectile and the instantaneous flow-field that are unique for the projectile configuration and for the inviscid and viscous simulations. It can be observed in Figs. 11a and 11b that, from state "6" onwards the acceleration and $C_d$ curves of both the projectile configurations are completely different. This can be attributed to the effect of the configuration as well as the effect of viscosity which plays a major role in determining the complex projectile-wave interactions in the highly transient subsonic flow fields.

## IV. Conclusions

A computational analysis was performed using a moving grid method to study the launch dynamics of supersonic projectiles. Both inviscid and viscous simulations have been carried out to analyze the flow field. It is revealed that the flow field contains many complicated unsteady flow structures that significantly affect the projectile aerodynamic characteristics. The results of the study are summarized below.

1. Two major phenomena which affect the flow field and the projectile aerodynamic characteristics are the projectile-secondary shockwave interaction, and the projectile overtaking process. It is found that the projectile configuration significantly affects its aerodynamic characteristics during its interaction with secondary shock wave in the unsteady flow field. At higher Mach numbers the interactions were observed to be stronger.

2. The overtaking process is always supersonic in nature owing to the attenuation of the blast wave Mach number. Neither the projectile configuration nor the viscosity affects the projectile aerodynamic characteristics to a great extent during the overtaking process for small blast wave Mach numbers, except in the impossible overtaking regime. But, at higher Mach numbers, the unsteady drag coefficient of the projectile changes drastically during the overtaking process. The change in drag coefficient during the overtaking process may seem to affect the overall unsteady drag coefficient of the projectile and hence will have some effect on the exterior ballistics of the projectile. This may obviously be more significant in the cylindrical projectile. However, the projectile configuration does not seem to decide the overtaking criteria regardless of the projectile Mach number.

3. The viscosity plays a major role in determining the flow structures in the flow fields. The presence of viscosity leads to vorticity diffusion. As a result of this, the shock patterns in the secondary blast wave are seen to be considerably weakened. It is also seen that the viscosity and projectile configuration play a major role in deciding the aerodynamic characteristics of the projectile in the impossible overtaking regime.